# Explainable Anatomy-Guided AI for Prostate MRI: Foundation Models and In Silico Clinical Trials for Virtual Biopsy-based Risk Assessment

Danial Khan[1, 6*], Zohaib Salahuddin[1*], Yumeng Zhang[1], Sheng Kuang[1], Shruti Atul Mali[1], Henry C. Woodruff[1, 2], Sina Amirrajab[1], Rachel Cavill[6], Eduardo Ibor-Crespo [3], Ana Jimenez-Pastor[3], Adrian Galiana- Bordera[4], Paula Jimenez Gomez[4], Luis Marti-Bonmati [4,5], Philippe Lambin [1, 2]

[1] Department of Precision Medicine, GROW - Research Institute for Oncology and Reproduction, Maastricht University, 6220 MD Maastricht, The Netherlands

[2] Department of Radiology and Nuclear Medicine, GROW - Research Institute for Oncology and Reproduction, Maastricht University, Medical Center+, 6229 HX Maastricht, The Netherlands

[3] Research & Frontiers in AI Department, Quantitative Imaging Biomarkers in Medicine, Quibim SL, Valencia, Spain

[4] Biomedical Imaging Research Group, La Fe Health Research Institute, Valencia, Spain

[5] Medical Imaging Department, La Fe University and Polytechnic Hospital, Valencia, Spain

[6] Department of Advanced Computing Sciences, Maastricht University, Maastricht, The Netherlands

* These authors contributed equally as first authors.

# Abstract

## Purpose:

To develop and validate a fully automated, anatomically guided deep-learning pipeline that combines foundation models with counterfactual explainability for prostate-cancer (PCa) risk stratification on routine magnetic-resonance imaging (MRI).

## Methods:

The pipeline comprises (i) an nnU-Net–based module that segments the prostate gland and its zonal anatomy on axial T2-weighted MRI; (ii) a classification module that fine-tunes the Universal Medical Pre-Trained (UMedPT) Swin-Transformer foundation model on 3-D image patches, optionally augmented with gland or zonal priors and with clinical variables; and (iii) a variational-autoencoder/generative-adversarial-network (VAE-GAN) framework that generates counterfactual examples to localise image regions driving the model's decisions. Development used 1,500 cases from PI-CAI for segmentation and 617 biparametric MRI examinations with clinical metadata from the CHAIMELEON challenge for classification (70 % train, 10 % validation, 20 % test). Performance was reported with Dice similarity coefficient (DSC) for segmentation and area under the ROC curve (AUC), balanced accuracy, F1 score and a composite CHAIMELEON score for classification. Clinical utility was assessed in a paired, multi-centre in-silico trial in which 20 clinicians interpreted the 125-case test set with and without AI support after a 60-day wash-out.

## Results:

Gland, peripheral-zone and transition-zone segmentations achieved mean DSCs of 0.95 ± 0.12, 0.94 ± 0.08 and 0.92 ± 0.21, respectively. Incorporating gland priors boosted the foundation model's AUC from 0.69 to 0.72, and a three-scale ensemble (patch sizes 160–224) obtained the best test performance (AUC = 0.79; composite score = 0.76), surpassing the 2024 CHAIMELEON challenge winners. Counterfactual heat-maps consistently highlighted lesion-containing regions within the segmented gland, providing intuitive, voxel-level explanations of risk predictions. In the prospective *in silico* trial, AI assistance increased mean diagnostic accuracy from 0.72 to 0.77 and Cohen's κ from 0.43 to 0.53, while cutting average review time per case from 5.3 min to 3.1 min (≈ 40 % gain).

## Conclusion:

Anatomy-aware foundation models enriched with gland priors and counterfactual explanations deliver accurate, transparent and time-saving PCa risk stratification on standard MRI, supporting their integration as virtual biopsies in clinical workflows.

# Introduction

Prostate cancer (PCa) is among the most prevalent malignancies affecting men, particularly those over 50 years of age[1]. It is the second most commonly diagnosed cancer in men and the sixth leading cause of cancer-related mortality globally[2]. In 2018, an estimated 1.28 million new cases and 359,000 deaths were reported. As global populations continue to age, these numbers are expected to nearly double by 2040, reaching 2.3 million cases and 740,000 deaths[2].

The prostate is divided into four anatomical zones: the peripheral zone (PZ), transition zone (TZ), central zone (CZ), and anterior fibromuscular stroma (AFS). The PZ is the most common site for cancerous growths, followed by the TZ. The CZ and AFS are rarely associated with PCa[3,4]. With age, the gland often enlarges due to benign prostatic hyperplasia (BPH), affecting primarely the TZ. By the age of 75, approximately 75% of men show signs of BPH[3], which can alter the morphology of the prostate and complicate cancer detection.

One of the main challenges in PCa diagnosis is its broad clinical spectrum—from indolent tumors that may never progress to life-threatening metastases[5]. Early detection and accurate grading are critical for guiding treatment and improving survival outcomes[6]. Traditional screening methods, such as prostate-specific antigen (PSA) testing and digital rectal exams (DRE), are effective for early detection but suffer from limited specificity, leading to overdiagnosis and unnecessary biopsies. Definitive diagnosis typically requires histopathological examination of biopsy samples obtained via transrectal or transperineal approaches, which carry procedural risks[7].

To assess cancer aggressiveness, pathologists use the Gleason Score (GS) and the Gleason Grade Group (GGG), which classify tumors based on glandular patterns to predict prostate cancer behavior and guide treatment[8]. The GGG system simplifies the GS into five clinically actionable groups. In our study, we further stratify cases into low-risk (GGG 1–2) and high-risk (GGG 3–5) categories, a division supported by clinical evidence indicating poorer outcomes and higher tumor volumes in high-risk groups[9].

Magnetic resonance imaging (MRI), especially multiparametric (mpMRI) and biparametric (bpMRI), has become an essential tool in PCa detection. mpMRI integrates T2-weighted (T2w), diffusion-weighted imaging (DWI), and dynamic contrast-enhanced (DCE) sequences, while bpMRI omits DCE for greater efficiency and lower cost[10]. These techniques enhance lesion visibility and risk assessment, particularly when paired with PSA and clinical data[11]. However, interpreting MRI requires specialized expertise and suffers from inter-observer variability, which hinders consistency and scalability in clinical settings[12].

Radiomics is a quantitative approach to medical image analysis that can be broadly categorized into handcrafted radiomics and deep learning–based radiomics[13]. Handcrafted radiomics involves the extraction of engineered features derived from predefined mathematical formulas[13]. In contrast, deep learning, a subset of machine learning, automatically learns to extract relevant features from images using complex multilayered architectures. Deep learning techniques have been widely applied to prostate cancer (PCa) detection and risk stratification from MRI [14,15].

Foundation models are pre-trained, large deep neural networks that can be adapted to a wide range of downstream tasks with minimal task-specific training[16]. Foundation models are built upon pre-trained encoders trained on large-scale medical imaging datasets[16]. These models differ significantly from conventional approaches in two key ways. First, they are trained on large-scale medical imaging datasets, enabling the learning of robust and generalizable image features. Second, they leverage self-supervised learning techniques or are trained across multiple supervised tasks simultaneously, allowing them to capture a broad spectrum of image representations. These learned features are task-agnostic and can be effectively transferred to a wide range of downstream applications. Universal biomedical pretrained model (UMedPT) is an advanced model trained using 17 shared tasks sourced from 15 publicly available datasets, covering diverse biomedical imaging modalities such as X-rays, MRIs, Computed Tomography (CT), microscopy, and histopathology images[17]. The training tasks span a variety of diseases, including brain tumors, lung cancer, prostate cancer, and COVID-19, and involve segmentation, classification, and detection. The model employs a multi-task learning framework, where shared blocks are trained across all tasks while task-specific heads like classification heads are fine-tuned for their respective objectives.

Despite the promising performance of deep learning methods in medical image analysis tasks, a major barrier to their adoption in clinical workflows is the lack of interpretability. These models function as "black boxes" that make it difficult for clinicians to trust or understand the rationale behind predictions [18,19]. To address this limitation, explainability techniques such as heatmaps and counterfactual examples have emerged. Counterfactual explanations go a step further than traditional heatmaps by illustrating how changes in the most influential image regions affect the model's predictions[20]. These methods visualize which features contribute most to a prediction to help in bridging the gap between AI decision-making and clinical reasoning[18].

In this study, we present a novel, fully automated pipeline for prostate cancer (PCa) risk classification from MRI, leveraging foundation models augmented with anatomical priors and counterfactual explainability. The proposed pipeline consists of three primary components: (1) a region-of-interest (ROI) segmentation module based on nnU-Net, designed to extract clinically relevant anatomical structures; (2) a classification module that demonstrates the superior performance of foundation models relative to conventional convolutional neural networks (CNNs), particularly when informed by anatomical context; and (3) an explainability framework utilizing variational autoencoder–generative adversarial networks (VAE-GANs) to generate counterfactual examples and heatmaps, thereby facilitating intuitive interpretation of model predictions. To evaluate the clinical utility of the proposed system, a multi-center prospective *in silico* trial was conducted in which clinicians performed risk classification on independent new cases with and without access to the AI model.

# Methods

## Datasets

Two datasets were utilized in this study to develop and evaluate the proposed pipeline for automated region-of-interest (ROI) detection and prostate cancer (PCa) risk classification. A brief overview of each dataset is provided below.

### PI-CAI Dataset

The PI-CAI dataset is a large-scale, multi-institutional resource designed to facilitate the development of AI algorithms for PCa lesion detection[12]. It includes biparametric MRI (bpMRI) scans with T2-weighted (transversal, sagittal, and coronal), diffusion-weighted imaging (DWI), and apparent diffusion coefficient (ADC) maps, alongside expert annotations and biopsy-confirmed clinical outcomes. A publicly available subset of 1,500 cases from three centers provides comprehensive prostate gland and zonal (PZ, TZ) annotations, as well as clinical variables such as age, PSA levels, PSA density, and Gleason scores.

### CHAIMELEON Dataset

The CHAIMELEON dataset is a large-scale imaging repository, curated to support AI development for various cancer types[21]. In this study, only the prostate cancer (PCa) subset was used, made available during the championship phase of the CHAIMELEON challenge. The dataset includes multiple MRI sequences, with T2-weighted images consistently available across all cases.

A total of 636 cases were included, stratified into a training set (446 cases, 70%), validation set (63 cases, 10%), and test set (127 cases, 20%), according to the risk group. Each case was annotated with clinical metadata including age, PSA levels, Gleason score (GS), ISUP grade group (GGG), and a binary risk label. High-risk cases were defined by a GGG of 3 or higher. The training set included 324 high-risk and 122 low-risk cases; the validation set included 17 high-risk and 46 low-risk cases; and the test set included 36 high-risk and 91 low-risk cases. The median patient age was 68 years (IQR: 10), and the median PSA level was 7.5 ng/mL (IQR: 5.37).

The dataset used in this study was collected from two medical centers: Centre Hospitalier Universitaire (CHU) d'Angers in France, which contributed 84 cases, and the Hospital Universitari i Politècnic La Fe in Valencia, Spain, which contributed 551 cases. MRI scans were acquired using scanners from multiple vendors, including Philips, Siemens Healthineers, Toshiba Medical Systems, and GE Healthcare, providing a heterogeneous imaging dataset representative of real-world clinical practice.

The dataset was filtered based on the availability of T2-weighted (T2w) MRI sequences, axial orientation, a slice thickness of ≤7 mm, and an in-plane resolution of ≤0.7 mm. Further refinement was performed using gland segmentations to crop the images around the relevant anatomical regions. The final dataset comprised 429 cases for training, 63 cases for validation, and 125 cases for testing. Five test cases were manually corrected to ensure accurate transversal orientation consistency. Model development followed a standard protocol: training was conducted on the training set, and performance was monitored on the validation set to select the best-performing models. Final evaluation was conducted once on the held-out test set using the best model identified through validation performance.

## Proposed Fully Automatic Pipeline

The fully automated pipeline comprises three main components, as illustrated in Figure 1: the region-of-interest (ROI) pipeline, the classification pipeline, and the explainability module. The ROI pipeline is responsible for automatically detecting and segmenting the prostate gland and its anatomical zones from T2w MRI scans. It prepares the input for the classification stage by extracting the relevant ROIs from the imaging data. The classification pipeline predicts prostate cancer risk based on the segmented ROIs and associated clinical features. The explainability module interprets the model's predictions by identifying and visualizing the key features influencing the classification outcome. The entire pipeline is designed for end-to-end automation, enabling seamless integration of MRI and clinical data for risk stratification of prostate cancer.

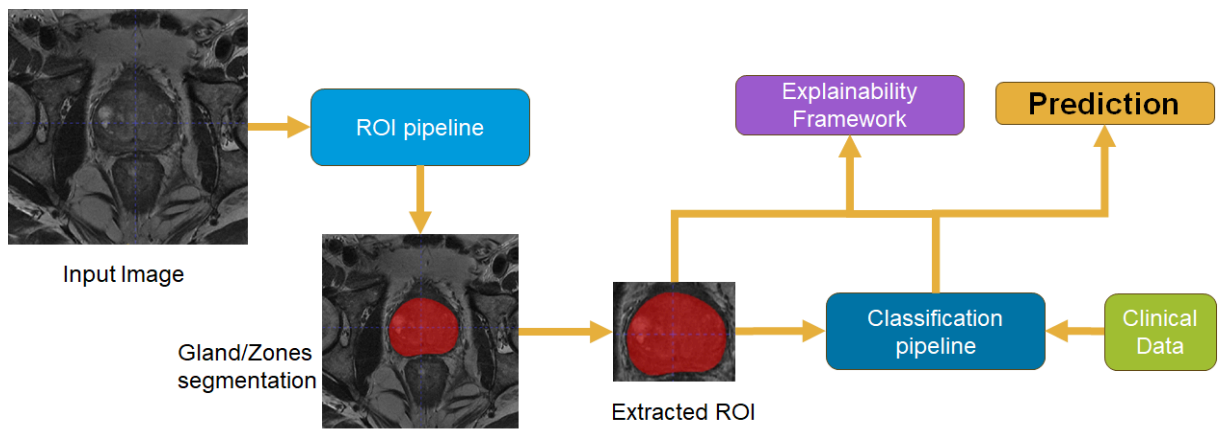

**Figure 1**: Overview of the fully automated pipeline for prostate cancer risk prediction. The pipeline includes three main components: (1) an ROI pipeline for segmenting the prostate gland and zones from T2w MRI, (2) a classification pipeline for predicting cancer risk using segmented ROIs and clinical data, and (3) an explainability module for interpreting model predictions.

## Region of Interest Delineation

The region of interest (ROI) pipeline was developed to automatically segment the prostate gland and its anatomical zones (peripheral zone [PZ] and transition zone [TZ]) from axial T2w MRI. It is based on nnU-NetV2, a state-of-the-art framework for medical image segmentation[22,23]. Two separate models were trained using the PI-CAI dataset: one for whole-gland segmentation and another for zonal segmentation. Both models shared identical configurations, differing only in the training labels.

Images were converted from MHA to NIfTI format and organized following nnU-Net's required structure. The default preprocessing pipeline was applied, including Z-score normalization and resampling to median spacing (0.5, 0.5, 3 mm). The 3D full-resolution configuration with a Residual Encoder U-Net architecture was employed, constrained to 24

GB GPU memory. The dataset was split 70/30 for training and testing, and models were trained using stratified 5-fold cross-validation with a batch size of 4 and patch size of (320, 320, 20) for 500 epochs on a single NVIDIA RTX 3090 GPU. Final predictions on the test set were obtained using an ensemble of the five folds with test-time augmentations.

Model performance was evaluated using the Dice Similarity Coefficient (DSC) on the held-out 30% test split of the PI-CAI dataset. The trained models were then applied to the CHAIMELEON dataset for gland and zonal segmentation. Post-processing using connected component filtering retained the largest structure in each segmentation mask, reducing noise. These refined segmentations were used to crop the images around the relevant anatomical regions, preparing inputs for the classification pipeline.

## Pre-processing

To prepare the data for the classification pipeline, T2w images from the CHAIMELEON dataset were first converted from their native DICOM format to NIfTI. Prostate gland and zonal segmentations, generated by the ROI pipeline, were then extracted. Both images and segmentation masks were resampled to the dataset's median spacing of (0.3125, 0.3125, 3 mm) using B-spline interpolation for images and nearest neighbor interpolation for masks. Cropping was performed around the region of interest, centered on the prostate gland. To capture volumetric information at multiple scales, three distinct patch sizes were extracted from the images and corresponding masks: (224, 224, 28), (192, 192, 24), and (160, 160, 20).

## Classification Pipeline

### CNN-Based Architecture

The CNN-based classification model was implemented in PyTorch using components from the MONAI and nnU-Net frameworks. The architecture was based on a modified SeResNet-34 and trained on 3D image patches of size (224, 224, 28). Anisotropic convolutional kernels and strides were employed to handle the non-uniform spatial resolution of medical imaging data. Data augmentation during training included affine transformations (scaling, rotation, shearing, flipping), intensity-based perturbations (contrast, brightness, gamma correction), Gaussian noise, and MRI-specific artifacts such as Gibbs and bias field noise. Detail on augmentation parameters is available in Appendix A.1.

The model incorporated a redesigned stem with a kernel size of (3×3×1), Leaky ReLU activations throughout, and anisotropic strides configured as (2,2,1) for the first block, (2,2,2) for the middle blocks, and (2,2,1) for the final block. Feature maps progressed through dimensions 32, 64, 128, 256, and 320. A 3D Adaptive Max Pooling layer reduced the spatial dimension to 4×4×4 before passing the output to a multilayer perceptron (MLP) classifier. The MLP included layers with 20,480, 4,096, and 1,024 neurons, followed by a sigmoid-activated output neuron for binary classification.

Training was conducted using alpha-balanced focal loss ($\alpha = 0.8$, $\gamma = 2.0$) and optimized with SGD (learning rate = $1 \times 10^{-3}$, weight decay = $1 \times 10^{-6}$). A cosine annealing learning rate scheduler was employed over 250 epochs with a batch size of 6. Half-precision training with gradient scaling and clipping was used to maintain numerical stability. Hyperparameter configurations are detailed in Appendix A.2.

## Foundation Model - UMedPT

The Universal Biomedical Pretrained model (UMedPT) foundation model[17], based on a Swin Transformer encoder, was fine-tuned for 3D classification by leveraging its pretrained components: a 2D encoder, a feature squeezer, and a custom classification pipeline.

3D volumes were decomposed into 2D slices and passed through the frozen encoder and squeezer modules to obtain compact slice-level embeddings (final vector size: 512). These were then aggregated by a trainable grouper module that learned to combine slice-level features into a single volume-level embedding. The final classification was performed using a fully connected head.

The model was trained on the largest patch size (224, 224, 28), with intensity values normalized using percentile-based Min-Max normalization. The training utilized weighted binary cross-entropy loss (positive class weight = 2.342) and the AdamW optimizer (LR = 5e-4, weight decay = 1e-4). Half-precision training was enabled without gradient scaling/clipping. Cosine annealing scheduling was used over 200 epochs with an effective batch size of 256 (batch size = 8, gradient accumulation over 32 steps). Training was conducted on a single NVIDIA A30 GPU (12 GB). Hyperparameters are listed in Appendix A.3.

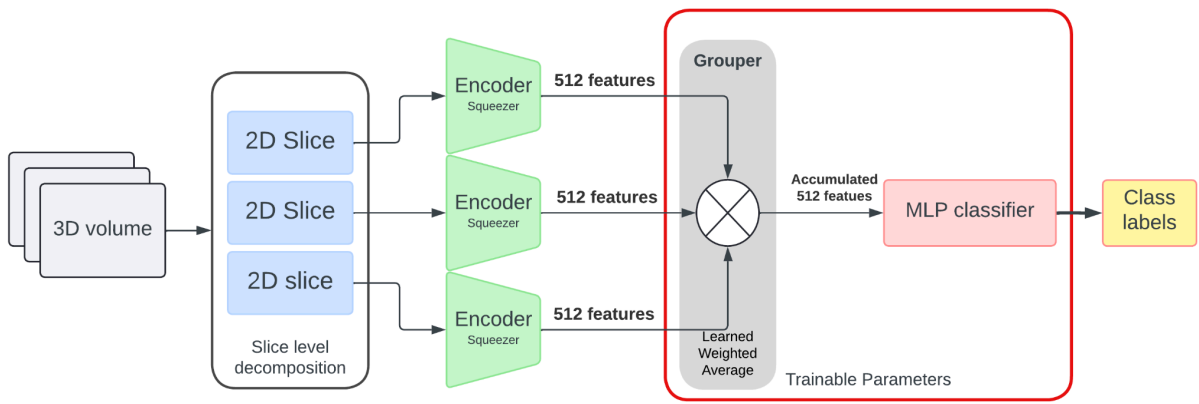

**Figure 2:** UMedPT architecture fine-tuning for a 3D classification task. The process starts with the decomposition of 3D images into 2D slices, which are then passed through the encoder to extract features. The squeezer compresses the features and reduces dimensionality. The grouper combines the features across all slices of a single volume and passes it to the classification head which predicts the class label.

## Enhancing the Foundation Model with Supplementary Information

To improve the predictive performance of the foundation model, we systematically explored the integration of supplementary information into the classification pipeline. Three types of additional data were investigated: clinical features, anatomical priors, and modified imaging inputs. These enhancements were incorporated into the foundation model architecture, and their impact was evaluated on both the validation and test sets.

### Clinical Feature Integration

Patient-level clinical data provided in the CHAIMELEON dataset, including age and prostate-specific antigen (PSA) levels, were used to enrich the model input. In addition, prostate volume (in cc) was estimated using the gland segmentation model developed in the ROI pipeline. This enabled the calculation of PSA density, which was included as a more informative biomarker compared to raw PSA values. Prior studies have shown that PSA density better accounts for anatomical variability across individuals[24].

All clinical features were standardized and concatenated with the deep feature vectors extracted from the imaging pipeline. These were then passed to the MLP classifier within the foundation model.

### Incorporation of Anatomical Priors

To provide spatial context and anatomical cues, segmentation-derived priors representing the prostate gland and its internal zones were included in the model input. These priors were generated using segmentation models from the ROI pipeline and added as an additional channel alongside the imaging data.

To preserve anatomical consistency, intensity-based augmentations were excluded from the priors during training. However, spatial transformations such as scaling and rotation were retained to ensure the model could learn robust spatial representations.

### Multi-Channel Imaging Input Strategy

To evaluate the combined impact of imaging, anatomical, and clinical information, we implemented a multi-channel input strategy. In the foundation model, the T2w sequence was duplicated to form two channels, and the anatomical prior was included as a third channel. This configuration allowed the model to simultaneously process both imaging detail and spatial anatomical context. Additionally, clinical features were appended to the learned feature embeddings before classification. A grid search was performed to identify the most effective configuration of input features and hyperparameters.

### Patch Size Sensitivity Analysis

To assess the influence of spatial resolution on model performance, the foundation model was retrained using alternative patch sizes of (192, 192, 24), and (160, 160, 20). The initial configuration used patches of size (224, 224, 28). All other training parameters and strategies remained consistent. These experiments were designed to evaluate how different patch sizes affect the model's ability to capture relevant imaging features. To evaluate whether multiscale patches leveraging foundation models can enhance performance by capturing features at different spatial resolutions, we performed an ensemble by averaging the predicted probabilities from models from: (160, 160, 20), (192, 192, 24) and (224, 224, 28). Figure 3 shows the ensemble configuration from multiscale patches.

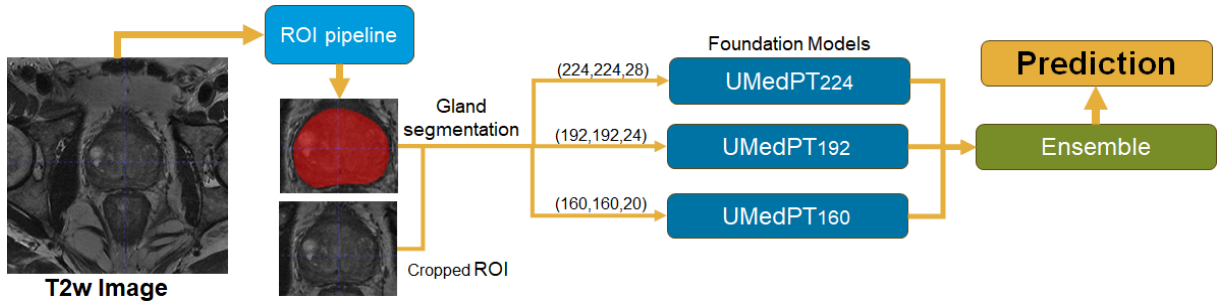

**Figure 3.** *The ensemble pipeline is based on UMedPT models trained with different patch sizes and incorporating gland priors. The final prediction is obtained by averaging the predicted probabilities across the models.*

# Explainability

Counterfactual explanations involve generating modified versions of an input image that lead to different prediction outcomes[25]. By exploring these "what-if" scenarios, we can identify the minimal changes required to alter the model's decision, offering insights into which image regions most strongly influence classification. These changes are visualized through saliency maps or heat maps that highlight critical anatomical zones. While counterfactual methods are commonly applied to 2D images due to computational constraints, we implemented a 3D VAE-GAN framework to support volumetric interpretability. This framework, adapted from the MONAI library, learns meaningful representations of input volumes. The gradients from the classifier's output with respect to the latent space are used to perturb latent vectors and generate counterfactual samples. These synthetic images are fed back through the model to assess prediction changes, and the resulting image differences are visualized as heatmaps, pinpointing influential regions.

## VAE-GAN Framework

Variational Autoencoder-Generative Adversarial Network (VAE-GAN) combines the representational power of Variational Autoencoders (VAEs) with the generative capabilities of Generative Adversarial Networks (GANs). Figure 4 shows the VAE-GAN framework. The VAE's decoder functions as the GAN generator, while the GAN discriminator distinguishes between real and generated images. The VAE encoder maps an input image $x$ to a latent representation $z$, which is decoded to reconstruct the image $\bar{x}$. The discriminator learns to differentiate between $x$ and $\bar{x}$, assigning a likelihood score.

Training involves the optimization of the following total loss:

$$Loss_{Total} = L_{reconstruction} + w_{KL} \cdot L_{KL} + w_{perceptual} \cdot L_{perceptual} + w_{adversarial} \cdot L_{adversarial}$$

Where:

- *Reconstruction loss* measures pixel-level difference between input image $x$ and reconstructed image $\bar{x}$:

$$L_{reconstruction} = \frac{1}{N}\sum_{i=1}^{N}\left\|x_i - \bar{x}_i\right\|$$

- **KL divergence loss** measures divergence of the learnt latent space $q_\phi(z|x)$ from the prior distribution $p(z)$:

$$L_{KL} = D_{KL}(q_\phi(z|x) || p(z))$$

- **Perceptual loss** measures measures the difference between the high-level feature representations of the images utilizing a pre-trained neural network to generate the feature maps $\phi_l(x)$ and $\phi_l(\bar{x})$. $l$ is the layer index, M is the number of elements in the layer, and L is the number of layers in the neural network:

$$L_{perceptual} = \sum_{l=1}^{L} \frac{1}{M_l} \left\| \phi_l(x) - \phi_l(\bar{x}) \right\|^2$$

- **Adversarial loss** measures the discriminator's $Disc$ performance in distinguishing between the real and generated images:

$$L_{adversarial} = log(Disc(x)) + log(1 - Disc(\bar{x}))$$

The loss weights were set as follows: $w_{KL} = 10^{-6}$, $w_{perceptual} = 10^{-3}$, and $w_{adversarial} - 10^{-2}$. The adversarial component was introduced after 10 warm-up epochs. The model was trained for 1000 epochs using a batch size of 4, with learning rates of $1 \times 10^{-4}$ for the autoencoder and $1 \times 10^{-5}$ for the discriminator, on a single NVIDIA V100 GPU with 32 GB memory. Early stopping with a patience of 200 epochs was used to prevent overfitting. Training was conducted on the CHAIMELEON dataset using the smallest patch size (160, 160, 20), and evaluation was performed on the test set. Reconstruction fidelity was assessed by comparing classification predictions on original $x$ and reconstructed images $\bar{x}$, ensuring minimal deviation. Cases with prediction differences less than 0.1 were selected as reference points for generating counterfactuals.

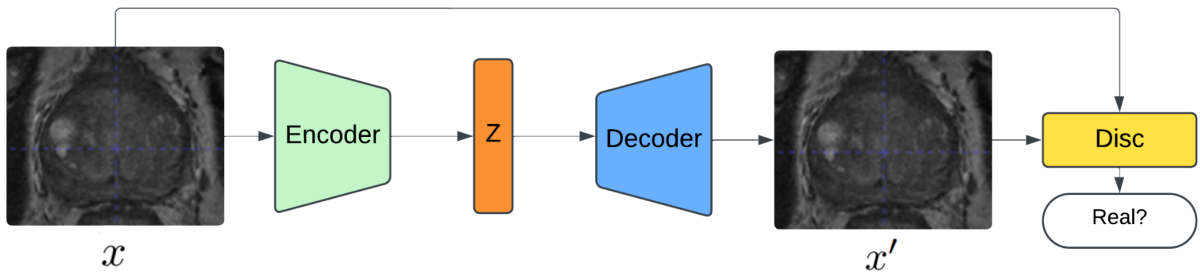

**Figure 4**: Illustration of the VAE-GAN framework. The architecture comprises two main components: a Variational Autoencoder (VAE) and a Generative Adversarial Network (GAN). The VAE encoder maps the input image $x$ to a latent representation $z$, which is then passed through the decoder to reconstruct the image $\bar{x}$. The GAN includes a discriminator that distinguishes between real and reconstructed images. Both the VAE and GAN components are trained concurrently to improve the realism and fidelity of the generated reconstructions.

## Counterfactual Explanations

Following training of the VAE-GAN model, counterfactual images were generated by perturbing the latent vector $z_{orig}$ using the gradient of the classifier's output with respect to the latent space:

$$x_{cf} = D\left(z_{orig} + \alpha \cdot \frac{\delta(\Sigma pred)}{\delta z_{orig}}\right)$$

Here, $\alpha$ is a step size that controls the magnitude of perturbation. This process was repeated iteratively to find the lower and upper bounds where prediction probabilities begin to shift significantly. Figure 5 shows the counterfactual explanation framework.

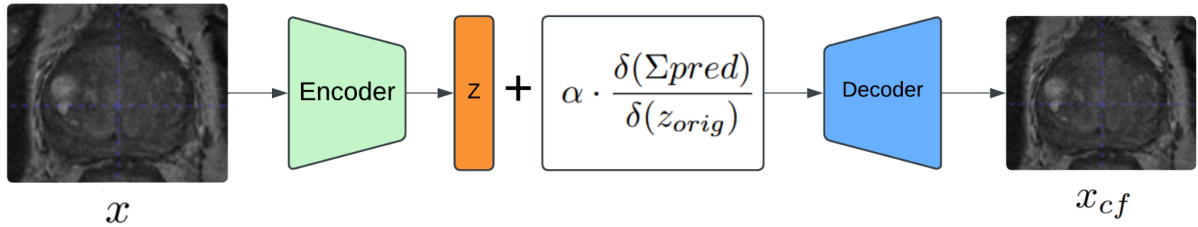

**Figure 5**: Illustration of the counterfactual explanation framework. The input image $x$ is passed through the encoder to generate a latent space representation $z$, which is used to reconstruct the original image. Counterfactual images are generated by perturbing this latent vector using gradients derived from the classification model's predictions.

Counterfactual samples were generated across a range of $\alpha$ values to explore model sensitivity. Heatmaps were computed by subtracting original from counterfactual images, highlighting regions most influential in changing model predictions. Two types of heatmaps were produced: one summarizing consistently modified regions across all counterfactuals, and another visualizing sequential differences between consecutive samples to capture dynamic decision shifts. These visualizations provided deeper insight into the model's decision-making process and highlighted the anatomical zones most critical to risk classification.

## Prospective *in silico* Trial

To evaluate the clinical utility of the proposed AI model for prostate cancer high-risk prediction, an in silico trial was conducted using the CHAIMELEON validation platform[1]. In the first phase of the trial, 20 clinicians participating in the prostate cancer use case were presented with the full test set of cases, including all available MRI images, along with relevant clinical information such as age and PSA levels. No AI assistance was provided at this stage. After a 60-day washout period to mitigate recall bias, the same clinicians re-evaluated the identical cases, this time with the addition of AI-generated high-risk predictions. Readers consisted of 19% of clinicians with less than 5 years of experience, 44% with 5 to 10 years, and 37% with more than 10 years of clinical experience. The performance across both phases was compared in terms of diagnostic accuracy, time taken per case, and inter-rater agreement measured using Cohen's kappa. This evaluation involved clinicians from 11 partner institutions, including the National Cancer Institute in Vilnius (Lithuania), the Local Health Unit of Santo António (Portugal), Charité – Universitätsmedizin Berlin (Germany), the Health Research Institute La Fe in Valencia (Spain), the University of Pisa (Italy), the Local Health Unit (Portugal), the Cancer Education and Research Foundation (Canada), Virgen del Rocío University Hospital in Seville (Spain),

---

[1] https://in-silico-validation.chaimeleon-eu.i3m.upv.es/

Osatek at Galdakao Hospital (Spain), Patient Specific Directions (United Kingdom), and the University of Messina (Italy).

## Evaluation Metrics

For the segmentation task, we used the DSC as the primary evaluation metric. The DSC measures the spatial overlap between the predicted and ground truth segmentations and is defined as follows: $DSC = \frac{2 \times |A \cap B|}{|A| + |B|}$, where A is the set of predicted segmentation pixels and B is the set of ground truth pixels. A DSC value of 1 indicates perfect overlap, while 0 indicates no overlap.

To evaluate model performance for the classification task, area under the receiver operating characteristic curve (AUC-ROC) was used, which quantifies the model's ability to discriminate between the two classes across a range of classification thresholds. The AUC-ROC score ranges from 0 to 1, with higher values indicating superior discriminative performance. However, given its sensitivity to class imbalance, additional complementary metrics were used to provide a more robust assessment. These included sensitivity (true positive rate), specificity (true negative rate), balanced accuracy (the average of sensitivity and specificity), and the F1 score (the harmonic mean of precision and recall), offering a balanced view of performance, especially in imbalanced datasets. To aggregate the evaluation into a single metric for ranking purposes, we computed a composite score using a weighted sum of the AUC-ROC, balanced accuracy, sensitivity, and specificity, as defined by the CHAIMELEON scoring protocol:

$$Score = 0.4 * AUCROC + 0.2 * Balanced\ Accuracy + 0.2 * Sensitivity + 0.2 * Specificity$$

Furthermore, Cohen's kappa was included to quantify the level of agreement for the in silico trial performance evaluation between the model's predictions and the ground truth, adjusted for chance agreement. Cohen's kappa ranges from −1 to 1, where 1 indicates perfect agreement, 0 represents agreement expected by chance, and negative values suggest disagreement.

Table 1: UMedPT 224 with Anatomical Priors and Clinical Variables, where G denotes the addition of the prostate gland mask, Z denotes the addition of prostate zone masks, and C denotes the addition of clinical variables.

| Model | Score | AUC | Balanced Accuracy | F1 Score | Sensitivity | Specificity |
|---|---|---|---|---|---|---|
| Challenge Winners | 0.669 | 0.716 | 0.637 | 0.479 | 0.472 | 0.802 |
| UMedPT 224 | 0.677 | 0.688 | 0.669 | 0.517 | 0.455 | **0.884** |
| UMedPT 224+G | **0.720** | 0.722 | **0.720** | **0.585** | **0.706** | 0.733 |
| UMedPT 224+Z | 0.670 | 0.728 | 0.632 | 0.471 | 0.485 | 0.779 |
| UMedPT 224+C | 0.662 | 0.656 | 0.666 | 0.520 | 0.588 | 0.744 |
| UMedPT 224+G+C | 0.697 | **0.727** | 0.677 | 0.535 | 0.677 | 0.678 |

Table 2: UMedPT Performance with Different Patch Sizes

| Model | Score | AUC | Balanced Accuracy | F1 Score | Sensitivity | Specificity |
|---|---|---|---|---|---|---|
| Challenge Winners | 0.669 | 0.716 | 0.637 | 0.479 | 0.472 | 0.802 |
| UMedPT 224 | 0.677 | 0.688 | 0.669 | 0.517 | 0.455 | **0.88** |
| UMedPT 224+G | 0.720 | 0.722 | 0.720 | 0.585 | 0.706 | 0.733 |
| UMedPT 192+G | 0.750 | 0.767 | **0.738** | **0.605** | **0.765** | **0.711** |
| UMedPT 160+G | 0.746 | **0.793** | 0.714 | 0.578 | 0.706 | 0.722 |
| Ensemble - UMedPT + G (224, 192, 160) | **0.760** | **0.793** | **0.738** | **0.605** | **0.765** | **0.711** |

# Results

## Region of Interest Delineation

The gland segmentation model achieved a mean Dice Similarity Coefficient (DSC) of 0.95 ± 0.12, while the zonal segmentation model obtained a mean DSC of 0.94 ± 0.08 for the peripheral zone (PZ) and 0.92 ± 0.21 for the transition zone (TZ) on the PI-CAI test set. When applied to the CHAIMELEON dataset, minor manual corrections were required in five training cases, two validation cases, and four test cases. The model failed to generate a valid segmentation in only one case from the training set.

## Classification Results

The DL 224 model achieved a score of 0.67, with an AUC of 0.66, balanced accuracy of 0.65, F1 score of 0.51, sensitivity of 0.63, and specificity of 0.68. In comparison, UMedPT 224 yielded slightly superior overall performance, achieving a score of 0.68, AUC of 0.69, balanced accuracy of 0.67, F1 score of 0.52, sensitivity of 0.45, and the highest specificity among all models at 0.88. The CHAIMELEON challenge winners achieved a score of 0.67, an AUC of 0.72, balanced accuracy of 0.64, F1 score of 0.48, sensitivity of 0.47, and specificity of 0.80.

To assess the utility of incorporating additional information in the foundation model, we performed a series of experiments by integrating anatomical priors and clinical variables into the UMedPT 224 architecture as shown in Table 1. The inclusion of gland segmentation priors (UMedPT 224+G) led to the most significant improvement, with the model achieving a score of 0.72, AUC of 0.72, balanced accuracy of 0.72, F1 score of 0.59, and sensitivity of 0.71. In contrast, the inclusion of clinical variables alone (UMedPT 224+C) resulted in a decrease in overall performance (score: 0.66), indicating that these variables were not effectively leveraged by the model without further tuning. When both gland priors and clinical variables were combined (UMedPT 224+G+C), the model still outperformed the baseline, achieving a score of 0.70. The addition of zone priors (UMedPT 224+Z) yielded a high AUC of 0.73 but led to a drop in the overall score to 0.67.

We further explored the effect of input patch size on model performance as listed in Table 2. Using the UMedPT model with gland priors, we compared patch sizes of 160, 192, and 224. The UMedPT 192+G model outperformed all others with a score of 0.75, balanced accuracy of 0.74, and F1 score of 0.60. The UMedPT 160+G model achieved a slightly lower score of 0.75 but recorded the highest AUC of 0.79. The UMedPT 224+G model, while still superior to the baseline, achieved a lower score of 0.72. The ensemble model, created by averaging the predicted probabilities from the three best-performing UMedPT configurations with gland priors (patch sizes 224, 192, and 160), achieved the highest overall performance. This approach yielded an overall score of 0.76, with an AUC of 0.79 and balanced accuracy of 0.74.

# In Silico Trial Results

Figure 6 summarizes the performance outcomes from the in silico trial under three conditions: clinician-only assessment, AI-only predictions, and clinician assessment supported by AI. The diagnostic accuracy achieved by clinicians alone averaged 0.72, while the AI model alone yielded an accuracy of 0.75. When clinicians were assisted by AI predictions, accuracy improved to 0.77, demonstrating a meaningful enhancement in diagnostic performance when AI is used as a decision-support tool rather than in isolation.

Cohen's kappa also reflected this trend, increasing from 0.43 in the clinician-only phase to 0.53 with AI assistance, compared to 0.50 for the AI model alone. In terms of efficiency, the average time spent per case was 5.3 minutes for clinician-only review, reduced to 3.1 minutes when supported by AI. These results confirm that AI integration not only boosts diagnostic accuracy and agreement but also substantially reduces the time burden on clinicians, supporting its utility in clinical workflows.

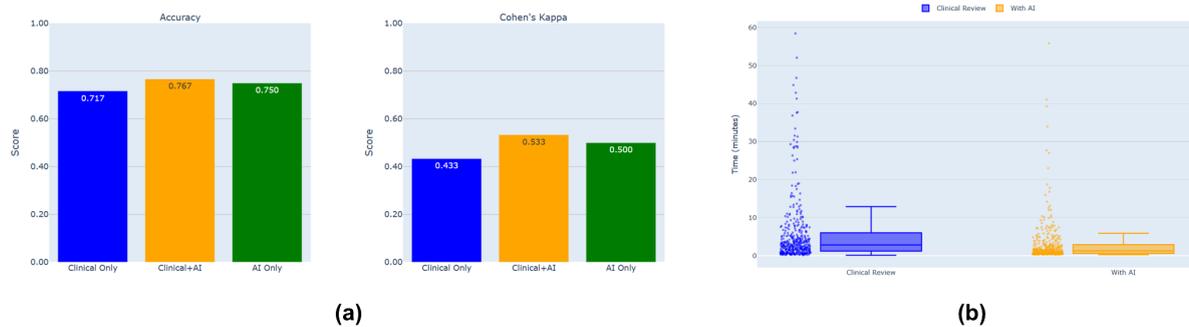

Figure 6: (a) Performance Comparison in terms of accuracy and Cohen's kappa for clinicians only, clinicians with AI assistance and AI alone. (b) Overall time distribution for clinical review with and without AI.

# Explainability

## VAE-GAN

To evaluate the impact of VAE-GAN–based reconstruction on the UMedPT model's predictions, the difference in probability scores between the original and reconstructed images was computed. Among the 125 test cases, 68 exhibited a difference of less than 0.1, indicating minimal effect on model outputs. The remaining 57 cases showed a greater deviation, with several outliers exceeding a difference of 0.2. Full details of this analysis are provided in Appendix A4. The subset of 68 cases with low prediction deviation was selected for generating counterfactual explanations.

## Counterfactual Explanations

The counterfactual explanations generated by the decoder of the VAE, trained within the VAE-GAN framework, are shown in Figure 7. These illustrate the effects of minor perturbations introduced into the latent space of input images. The resulting counterfactual images are displayed with their corresponding model prediction scores, revealing how predictions vary. The second row shows the absolute difference between the reference and counterfactual images, highlighting the most prominent areas of change. Notably, the model's prediction shifts from 0.26 to 0.53, marking a transition from low-risk to high-risk classification. The most significant changes are localized within the prostate gland, particularly around two lesion areas, which are clearly visible in the absolute difference maps.

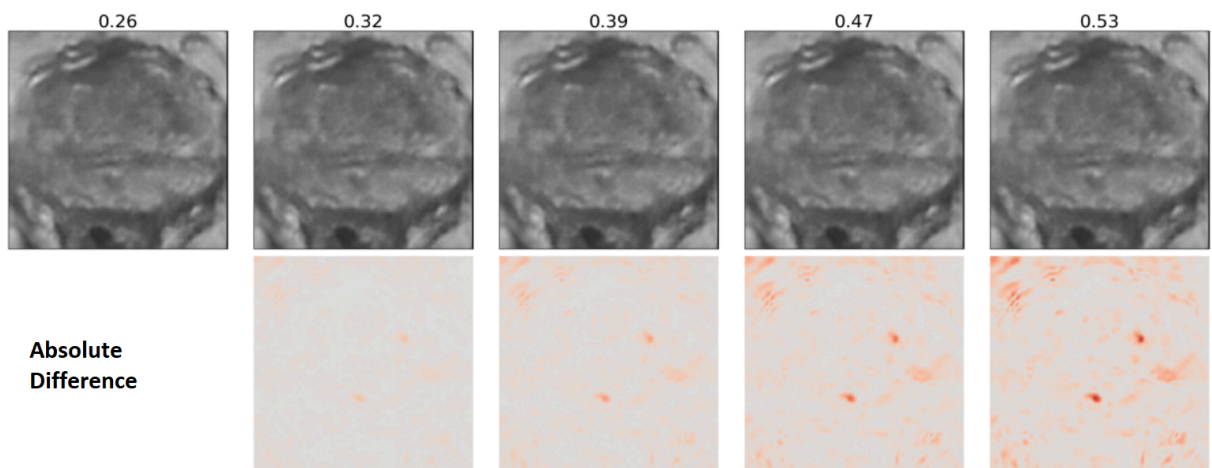

**Figure 7.** *Counterfactual explanations generated using the VAE-GAN. The first row presents the reference image alongside counterfactual images produced by perturbing the latent space of the input image. The model's predicted probability for each image is indicated above. The second row displays the absolute difference between the reference and counterfactual images, highlighting the most prominent regions of change.*

Figure 8 demonstrates the model's behavior in response to more substantial perturbations that lead to changes in image contrast. As the images progress from left to right, tissue homogeneity decreases and intensity contrasts become more pronounced. Comparing the outermost images with the central reference image shows that the model associates homogeneous tissue appearance with low-risk cancer (prediction of 0.24) and sharp intensity variations, resembling lesion patterns, with high-risk cancer (prediction of 0.73). The most noticeable contrast changes, aligned with the model's shifting predictions, are again concentrated within the prostate gland.

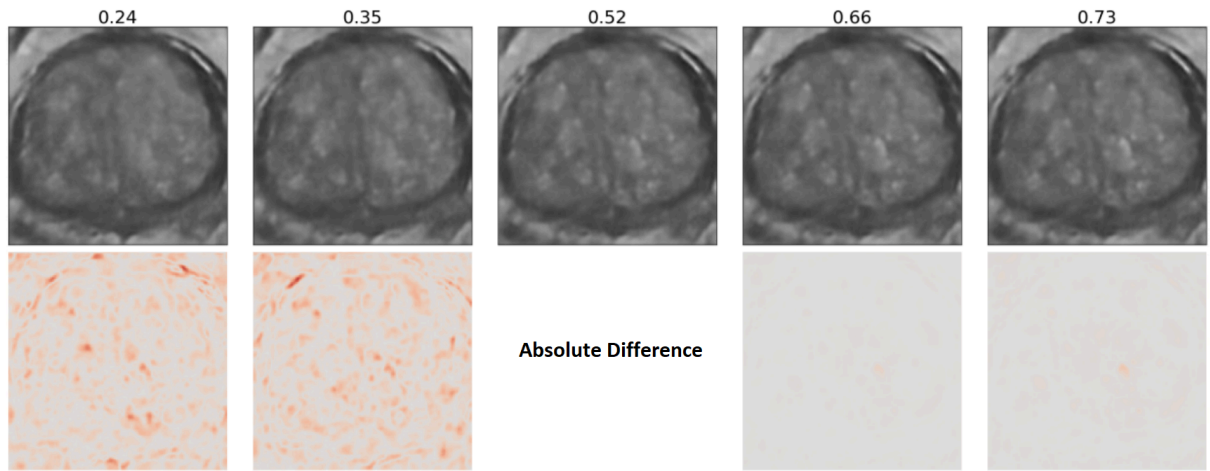

**Figure 8.** *Counterfactual explanations generated using the VAE-GAN framework. The middle column displays the reference image, with counterfactual images shown on either side. Model prediction probabilities are indicated above each image. The second row presents the absolute difference between the reference and counterfactual images, highlighting the most prominent regions of change.*

# Discussion

Our study presents a fully automated pipeline for prostate cancer detection and risk stratification, which first segments the prostate gland and then performs classification using a state-of-the-art foundation model enhanced with deep learning and explainability mechanisms. To improve model performance, anatomical priors were incorporated, resulting in a significant improvement over the challenge-winning approaches. The proposed method was validated in an international, multicenter prospective in silico trial involving 20 clinicians from 11 partner institutions. The results demonstrated that integrating AI support into clinical workflows not only enhances diagnostic accuracy and inter-rater agreement but also substantially reduces clinician workload, as reflected in reduced case review times. Additionally, a VAE-GAN framework was employed to learn the latent representations of test images, enabling the generation of counterfactual explanations by perturbing the latent space based on classification gradients. These counterfactuals were used to produce heatmaps, effectively visualizing the intensity and contrast changes that most influenced the model's predictions, thereby offering interpretable insights into the model's decision-making process.

The foundation model UMedPT demonstrated competitive performance with minimal task-specific tuning compared to standard deep learning models. Its ability to effectively leverage anatomical priors and clinical data can be attributed to its shared architecture, pretrained on diverse tasks, datasets, and modalities, allowing it to extract generalizable features applicable across downstream applications[17,26]. This broad pretraining minimizes the need for extensive retraining on large task-specific datasets while maintaining high performance.

Among the additional inputs tested, the inclusion of gland priors yielded the most substantial improvement. By explicitly directing the model's attention to the prostate gland, an anatomically relevant region for classification, the model's accuracy improved significantly. In contrast, the addition of clinical data and zonal priors had mixed effects. Age and PSA density did not provide sufficient predictive value for the model to learn effectively, resulting in increased false positives and reduced specificity, as reflected in a decline in AUC. This trend aligns with previous studies suggesting that such clinical indicators are more predictive in cases with small to medium sized glands [24]. Combining gland priors with clinical data showed modest improvements over the baseline model, though the performance gains were largely driven by the gland prior alone.

The variation of the input patch size also offered insights into model behavior. The UMedPT model using a patch size of 192 achieved the highest performance, followed by 160 and 224. AUC scores improved as patch size decreased, suggesting increased model confidence and improved discrimination. This pattern was consistent with models trained from scratch and may be attributed to better localization of fine-grained textural features and reduced background noise, particularly in cases with smaller prostate volumes. These observations were further supported by experiments showing that decreasing interpixel spacing negatively impacted performance, reinforcing the importance of scale and context in image resolution for effective risk stratification.

The analysis of classification performance for Gleason score (GS) 7 cases revealed that 17 out of 54 low-risk cases were misclassified, compared to only 5 out of 21 high-risk cases. This discrepancy highlights two key challenges. First, the limited contrast and overlapping tissue characteristics in T2-weighted MRI make it inherently difficult to differentiate between low-risk and high-risk prostate cancer, particularly in the presence of mixed primary and secondary Gleason patterns. Previous studies have recommended the inclusion of additional MRI sequences to improve diagnostic accuracy for GS 7 cases[27]. Second, the higher misclassification rate observed in low-risk cases is likely influenced by class imbalance in the dataset, which may bias the model toward overprediction of the high-risk class.

The paired, multi-centre in-silico study confirmed that the proposed AI system augments the clinical value expertise with increased performance and efficiency. When the 20 participating clinicians read the 125-case test set unaided, mean accuracy and inter-rater agreement were 0.72 and κ = 0.43, respectively. With AI support provided after a 60-day wash-out, the accuracy rose to 0.77 and κ to 0.53, while the average reading time fell from 5.3 min to 3.1 min per case (≈ 40 % gain in efficiency). Similar findings were also observed in recent multicentre reader-studies that reported an increase in sensitivity and ~50–60 % reductions in reading time when radiologists used AI-aided workstations for detecting clinically significant prostate cancer[28]. The findings support incorporating the pipeline as a real-time "second pair of eyes" to enable faster and more consistent prostate cancer risk stratification, while ultimately leaving the final clinical decision to human experts.

The reconstructions generated by the VAE GAN framework were used to evaluate their effect on the model's predictions. For the majority of cases, the probability scores remained within a threshold of 0.1, suggesting that the reconstructions did not significantly alter the model's outputs. However, in a subset of cases, prediction scores deviated notably, indicating that certain reconstructions introduced latent artifacts that influenced the classification. These inconsistencies suggest that while VAE GAN reconstructions are generally robust, further refinement, such as incorporating ROI-guided learning or enhanced regularization, may be necessary to ensure more stable latent representations.

Counterfactual explanations generated using the VAE GAN provided interpretability by highlighting the image regions that influenced the model's decisions. The corresponding heat maps revealed two key findings: first, the most prominent changes occurred within the prostate gland; second, the model was particularly sensitive to variations around glandular lesions. As the counterfactuals transitioned from low-risk to high-risk predictions, darkened lesions within the gland became increasingly evident. These changes were driven by decreased signal intensity within the lesion and/or increased intensity in the surrounding tissue. This pattern aligns with the PI RADS assessment criteria, in which high-grade tumors (scores 4 and 5) typically appear as low-intensity regions on T2 weighted images, in contrast to low-grade tumors (score 2)[29]. Overall, the counterfactuals provided valuable insights into the model's behavior and enhanced interpretability by localizing features critical to its risk stratification.

The Gleason grading system is widely known for its lack of consistent agreement among pathologists[30] and discrepancy of the Gleason score evaluated after prostatectomy (the gold standard). Biopsy based Gleason scoring accuracy is limited by sampling variation, often missing the tumor's most aggressive area, and by discrepancies between pathologists' grading. Even with multi core biopsies, 25 to 50% of prostate cancer cases require score adjustments after surgery, and pathologist discordance can reach up to 30%, which can complicate consistent prognostication [31]. The proposed pipeline can be seen as virtual biopsy, where this non-invasive alternative could complement or precede traditional methods, improving patient comfort, disease management, and potentially accelerating the diagnostic process.

The models developed in this study were trained on a dataset containing only cancerous cases, which may limit their generalizability to datasets containing a mix of cancerous and noncancerous cases. Additionally, since MRIs are qualitative and highly variable in acquisition protocols, domain adaptation has not yet been tested on external datasets. The models were trained exclusively on T2w sequences due to incomplete data for diffusion weighted imaging (DWI). As a result, apparent diffusion coefficient (ADC) maps, derived from DWI, could not be included. Their incorporation can further improve the model performance. Another limitation is that the classification task was based on thresholds for Gleason Score (GS) and Grade Group (GGG), yet various stratification protocols treat GS 7 (3+4) and GS 7 (4+3) differently. Any changes to these criteria would necessitate retraining the models. Lastly, the counterfactual explanations were not evaluated by expert radiologists to investigate if these explanations help in improving the trust of expert radiologists in the AI system. Future work will focus on including multiple MRI sequences such as DWI and ADC maps in the algorithm as well as validation of the counterfactuals explanations by expert radiologists.

# Conclusion

This study demonstrates that anatomically guided foundation models, enriched with gland-specific priors and counterfactual explainability, can provide fully automated, reliable, and interpretable risk stratification for prostate cancer using routine MRI. The integration of anatomical prior in terms of gland segmentation with T2w MRI sequence, the proposed pipeline based on UMedPT foundation model outperformed previous challenge-winning approaches. When ensembled across multiple resolutions, the model achieved an AUC of 0.79 and an overall score of 0.76 on the test set. Importantly, the VAE GAN-based counterfactual framework enabled localization of image regions that contributed to individual predictions, thereby enhancing transparency and bridging the gap between AI outputs and radiological interpretation. In a 20-reader, 11-site in silico trial, the integration of AI support led to a 5-point improvement in diagnostic accuracy, a 10-point increase in Cohen's kappa , and a 40% reduction in review time. While further validation on larger, multi-sequence datasets is needed, these findings underscore the potential of anatomy-aware foundation models as scalable and interpretable tools that can function as "virtual biopsies" to streamline prostate cancer management and support more consistent, data-driven clinical decision-making.

## Grants and funding

Authors acknowledge financial support from ERC advanced grant (ERC-ADG-2015 n° 694812 - Hypoximmuno),, ERC-2020-PoC: 957565-AUTO.DISTINCT. Authors also acknowledge financial support from the European Union's Horizon research and innovation programme under grant agreement: CHAIMELEON n° 952172 (main contributor), ImmunoSABR n° 733008, EuCanImage n° 952103, TRANSCAN Joint Transnational Call 2016 (JTC2016 CLEARLY n° UM 2017-8295),  IMI-OPTIMA n° 101034347, AIDAVA (HORIZON-HLTH-2021-TOOL-06) n°101057062, REALM (HORIZON-HLTH-2022-TOOL-11) n° 101095435, RADIOVAL (HORIZON-HLTH-2021-DISEASE-04-04) n°101057699 and EUCAIM (DIGITAL-2022-CLOUD-AI-02) n°101100633.

## Disclosures

Disclosures from the last 36 months within and outside the submitted work: none related to the current manuscript; outside of current manuscript: grants/sponsored research agreements from Radiomics SA, Convert Pharmaceuticals and LivingMed Biotech. He received a presenter fee (in cash or in kind) and/or reimbursement of travel costs/consultancy fee (in cash or in kind) from Radiomics SA, BHV & Roche. PL has shares in the companies Radiomics SA, Convert pharmaceuticals, Comunicare, LivingMed Biotech, BHV and Bactam. PL is co-inventor of two issued patents with royalties on radiomics (PCT/NL2014/050248 and PCT/NL2014/050728), licensed to Radiomics SA; one issued patent on mtDNA (PCT/EP2014/059089), licensed to ptTheragnostic/DNAmito; one non-issued patent on LSRT (PCT/ P126537PC00, US: 17802766), licensed to Varian; three non-patented inventions (softwares) licensed to ptTheragnostic/DNAmito, Radiomics SA and Health Innovation Ventures and two non-issued, non-licensed patents on Deep Learning-Radiomics (N2024482, N2024889). He confirms that none of the above entities were involved in the preparation of this paper.

# References

1.  Miller, K. D. *et al.* Cancer treatment and survivorship statistics, 2022. *CA: A Cancer Journal for Clinicians* **72**, 409–436 (2022).

2.  Culp, M. B., Soerjomataram, I., Efstathiou, J. A., Bray, F. & Jemal, A. Recent Global Patterns in Prostate Cancer Incidence and Mortality Rates. *Eur Urol* **77**, 38–52 (2020).

3.  Simon, R. M. *et al.* Does Prostate Size Predict the Development of Incident Lower Urinary Tract Symptoms in Men with Mild to No Current Symptoms? Results from the REDUCE Trial. *Eur Urol* **69**, 885–891 (2016).

4.  Cohen, R. J. *et al.* Central zone carcinoma of the prostate gland: a distinct tumor type with poor prognostic features. *J Urol* **179**, 1762–7; discussion 1767 (2008).

5.  Boschheidgen, M. *et al.* MRI grading for the prediction of prostate cancer aggressiveness. *Eur Radiol* **32**, 2351–2359 (2022).

6.  Van Poppel, H. *et al.* Prostate-specific Antigen Testing as Part of a Risk-Adapted Early Detection Strategy for Prostate Cancer: European Association of Urology Position and Recommendations for 2021. *Eur Urol* **80**, 703–711 (2021).

7.  Rosario, D. J. *et al.* Short term outcomes of prostate biopsy in men tested for cancer by prostate specific antigen: prospective evaluation within ProtecT study. *BMJ* **344**, d7894 (2012).

8.  Atallah, C., Toi, A. & van der Kwast, T. H. Gleason grade 5 prostate cancer: sub-patterns and prognosis. *Pathology* **53**, 3–11 (2021).

9.  Hashmi, A. A. *et al.* International Society of Urological Pathology (ISUP)-Grade Grouping in Prostatic Adenocarcinoma and its Prognostic Implications. *Cancer Invest* **40**, 211–218 (2022).

10. Palumbo, P. *et al.* Biparametric (bp) and multiparametric (mp) magnetic resonance imaging (MRI) approach to prostate cancer disease: a narrative review of current debate on dynamic contrast enhancement. *Gland Surg* **9**, 2235–2247 (2020).

11. Eldred-Evans, D. *et al.* Rethinking prostate cancer screening: could MRI be an

alternative screening test? *Nat Rev Urol* **17**, 526–539 (2020).

12. Saha, A. *et al.* Artificial intelligence and radiologists in prostate cancer detection on MRI (PI-CAI): an international, paired, non-inferiority, confirmatory study. *Lancet Oncol* **25**, 879–887 (2024).

13. Lambin, P. *et al.* Radiomics: the bridge between medical imaging and personalized medicine. *Nat. Rev. Clin. Oncol.* **14**, 749–762 (2017).

14. Mehralivand, S. *et al.* Deep learning-based artificial intelligence for prostate cancer detection at biparametric MRI. *Abdom Radiol (NY)* **47**, 1425–1434 (2022).

15. Zhao, L. *et al.* Predicting clinically significant prostate cancer with a deep learning approach: a multicentre retrospective study. *Eur J Nucl Med Mol Imaging* **50**, 727–741 (2023).

16. Moor, M. *et al.* Foundation models for generalist medical artificial intelligence. *Nature* **616**, 259–265 (2023).

17. Schäfer, R. *et al.* Overcoming data scarcity in biomedical imaging with a foundational multi-task model. *Nat Comput Sci* **4**, 495–509 (2024).

18. Singla, S., Eslami, M., Pollack, B., Wallace, S. & Batmanghelich, K. Explaining the black-box smoothly—A counterfactual approach. *Med. Image Anal.* **84**, 102721 (2023).

19. Salahuddin, Z. *et al.* Counterfactuals and Uncertainty-Based Explainable Paradigm for the Automated Detection and Segmentation of Renal Cysts in Computed Tomography Images: A Multi-Center Study. *arXiv [eess.IV]* (2024).

20. Fang, Y. *et al.* DiffExplainer: Unveiling Black Box Models Via Counterfactual Generation. *arXiv [cs.CV]* (2024).

21. CHAIMELEON Consortium. OpenChallenge Championship Training Dataset for Prostate Cancer. Preprint at https://doi.org/10.5281/ZENODO.11454910 (2024).

22. Isensee, F., Jaeger, P. F., Kohl, S. A. A., Petersen, J. & Maier-Hein, K. H. nnU-Net: a self-configuring method for deep learning-based biomedical image segmentation. *Nat. Methods* **18**, 203–211 (2021).

23. Isensee, F. *et al.* NnU-Net revisited: A call for rigorous validation in 3D medical image

segmentation. *arXiv [cs.CV]* (2024) doi:10.48550/ARXIV.2404.09556.

24. Omri, N. *et al.* Association between PSA density and pathologically significant prostate cancer: The impact of prostate volume. *Prostate* **80**, 1444–1449 (2020).

25. Salahuddin, Z., Woodruff, H. C., Chatterjee, A. & Lambin, P. Transparency of deep neural networks for medical image analysis: A review of interpretability methods. *Comput. Biol. Med.* **140**, 105111 (2022).

26. Nicke, T. *et al.* Tissue concepts: Supervised foundation models in computational pathology. *Comput Biol Med* **186**, 109621 (2025).

27. Krishna, S. *et al.* Comparison of Prostate Imaging Reporting and Data System versions 1 and 2 for the Detection of Peripheral Zone Gleason Score 3 + 4 = 7 Cancers. *AJR Am J Roentgenol* **209**, W365–W373 (2017).

28. Sun, Z. *et al.* A multicenter study of artificial intelligence-aided software for detecting visible clinically significant prostate cancer on mpMRI. *Insights into Imaging* **14**, 1–12 (2023).

29. Prostate Cancer - PI-RADS v2.1. https://radiologyassistant.nl/abdomen/prostate/prostate-cancer-pi-rads-v2-1.

30. Mun, Y., Paik, I., Shin, S.-J., Kwak, T.-Y. & Chang, H. Yet Another Automated Gleason Grading System (YAAGGS) by weakly supervised deep learning. *NPJ Digit Med* **4**, 99 (2021).

31. Shipitsin, M. *et al.* Identification of proteomic biomarkers predicting prostate cancer aggressiveness and lethality despite biopsy-sampling error. *British Journal of Cancer* **111**, 1201–1212 (2014).

# Appendix

## A1. Augmentation Parameters

**Table A1:** MONAI Augmentation Parameters

| Augmentation | Transform Name | Parameters |
|---|---|---|
| Zoom | RandZoomd | prob=0.2, min_zoom=0.9, max_zoom=1.1 |
| Affine Transformations | RandAffined | prob=0.5, rotate_range=(0,0,π/15), shear_range=(0.1,0.1,0.1), scale_range=(0.1,0.1,0.1) |
| Flip | RandFlipd | prob=0.2 |
| Gaussian Noise | RandGaussianNoised | prob=0.1, mean=0.0, std=0.1 |
| Gaussian Smoothing | RandGaussianSmoothd | prob=0.1, sigma_x=(0.5,1), sigma_y=(0.5,1), sigma_z=(0.5,1) |
| Intensity Scaling | RandScaleIntensityd | prob=0.2, factors=(0.8,1.2) |
| Contrast Adjustment | RandAdjustContrastd | prob=0.2, gamma=(0.8,1.2) |
| Bias Field | RandBiasFieldd | prob=0.1, coeff_range=(0.1,0.2) |
| Gibbs Noise | RandGibbsNoised | prob=0.1, alpha=(0.6,0.8) |

## A2. DL224 Grid Search

**Table A2:** Grid Search Parameters – DL224 Model

| Parameter | Values Explored |
|---|---|
| Number of Epochs | 150, 200, **250**, 275, 300 |
| Batch Size | 2, 4, **6** |
| Learning Rate | 1e-4, 5e-4, **1e-3**, 2e-3 |
| Optimizer | AdamW, SGD |
| Momentum (SGD) | 0.85, 0.9 |
| Weight Decay | 1e-3, 1e-4, 1e-5, **1e-6** |
| Loss Function | BCE, **Focal Loss** |
| Positive Weight for BCE | 2.3, 2.5, 2.699 |
| Alpha (Focal Loss) | 0.71, 0.7268, 0.75, **0.8** |
| Gamma (Focal Loss) | 2, 4 |
| LR Scheduler | ReduceLROnPlateau, **Cosine Annealing** |
| Strides in Stem | **(2,2,1)**, (1,1,1) |
| Final Feature Map Size | **320**, 512 |
| Residual Block Configuration | ResNet-18 (2,2,2,2), **ResNet-34 (3,4,6,3)** |

## A3. UMedPT Models Grid Search

Table A3: Grid Search Parameters – UMedPT Model

| Parameter | Values Explored | Best |
|---|---|---|
| Number of Epochs | **200**, 225, 250 | 200 |
| Gradient Accumulation | 16, **32** | 32 |
| Learning Rate | 1e-3, 1e-4, 4e-4, **5e-4**, 6e-4 | 5e-4 |
| Optimizer | **AdamW**, SGD | AdamW |
| Weight Decay | **1e-4**, 1e-5, 1e-6 | 1e-4 |
| Loss Function | **BCE**, Focal Loss | BCE |
| Positive Weight for BCE | 2.3, 2.5, **2.699** | 2.699 |

## A4. VAE's generation comparative analysis

The comparative analysis of the VAE's generation and its influence on the model's prediction is illustrated in Figure A.1. The histogram shows the distribution of absolute differences in probability scores generated using the original and reconstructed images by the model. The histogram indicates that most cases show a difference in probability scores of less than 0.1, and most probability scores are near that threshold. However, some cases show a significant difference in probability scores, reaching up to 0.2 and some outliers at 0.25. Out of this distribution, 68 cases show a difference in probability scores of less than 0.1, while 57 cases show a difference in probability scores of more than 0.1 for the whole test set.

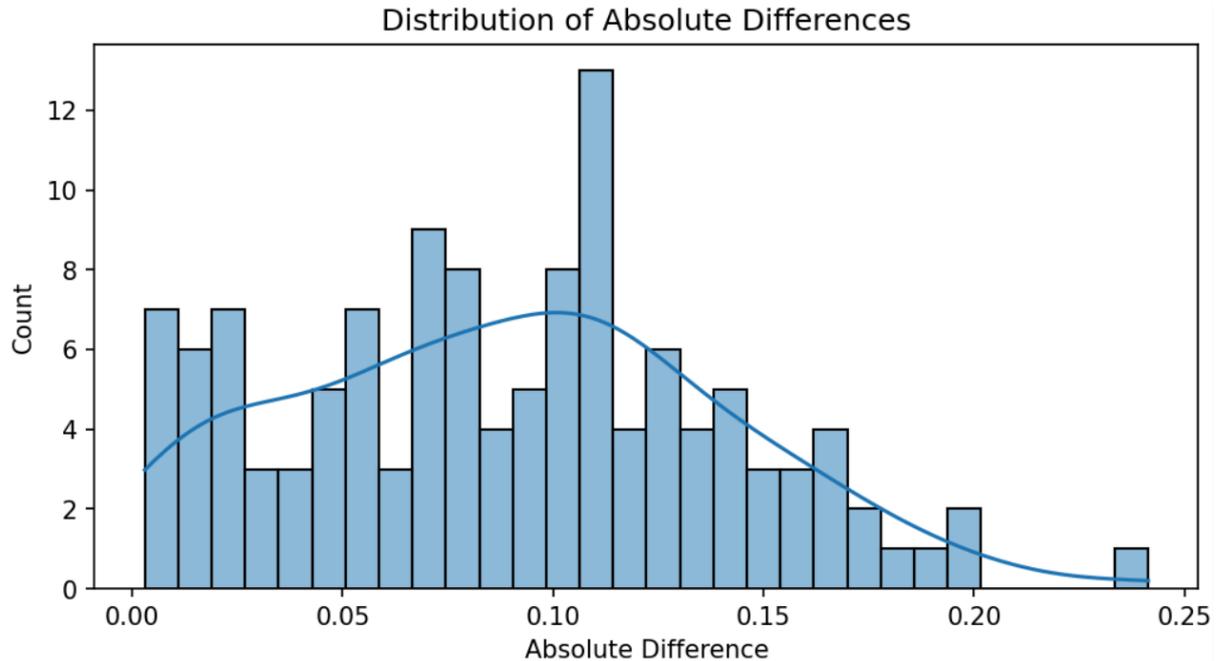

**Figure A.1:** Histogram presenting the absolute difference in probability scores generated using original and reconstructed images by UMedPT+G model. The x-axis represents the difference in probability scores, while the y-axis represents the number of cases. Most cases show a difference in probability scores of less than 0.1, indicating that the reconstructed images did not significantly impact the model's predictions.